\documentclass[article, twocolumn, superscriptaddress]{revtex4-1}
\usepackage[utf8]{inputenc}
\usepackage[T1]{fontenc}
\usepackage{amsmath, amssymb, mathtools, braket}
\usepackage{graphicx, subcaption}
\usepackage{pgfplots}
\pgfplotsset{compat=1.14}
\usepackage{csquotes}
\usepackage{listings}
\usepackage{color}
\usepackage[colorlinks=true, linkcolor=blue, citecolor=blue, urlcolor=blue]{hyperref}
\definecolor{mygreen}{rgb}{0,0.6,0}
\definecolor{mygray}{rgb}{0.5,0.5,0.5}
\definecolor{mymauve}{rgb}{0.58,0,0.82}
\begin{document}

\title{Quantum Simulation and Energy Estimation for Discretized Anharmonic Oscillator}

\author{Saurav Suman}
\email{sumansauravphy257@gmail.com}
\affiliation{National Institute of Technology Jamshedpur Adityapur, Jharkhand 831014}
\affiliation{CQST, Siksha O Anusandhan, Khandagiri, Bhubaneswar, 751030, Odisha, India}
\author{Bikash K. Behera}
\email{bikas.riki@gmail.com}
\affiliation{Bikash's Quantum (OPC) Private Limited, Mohanpur, 741246, West Bengal, India}
\author{Vivek Vyas}
\affiliation{Indian Institute of Information Technology Vadodara Gandhinagar, Gujarat, India}
\author{Prasanta K. Panigrahi}
\email{pprasanta@iiserkol.ac.in}
\affiliation{Department of Physical Sciences, Indian Institute of Science Education and Research Kolkata, Mohanpur 741246, West Bengal, India}
\affiliation{CQST, Siksha O Anusandhan, Khandagiri, Bhubaneswar, 751030, Odisha, India}
\begin{abstract}
Anharmonic potential quantum systems play a crucial role in physics as they provide a more realistic description of oscillatory phenomena, which often deviate from the idealized harmonic model. However, simulating such systems on classical computers is highly challenging due to nonlinear interactions, large state spaces, and the exponential scaling of memory and computational resources. In this work, quantum simulation is employed to model a quantum anharmonic oscillator (QAHO) using a 3-qubit system implemented on IBM's Quantum Experience platform. A quantum circuit with a filter-based design and Toffoli gates is constructed to track quantum state evolution, capturing key phenomena like quantum revival. The framework is further extended to n-qubit systems to enhance resolution and scalability. For energy estimation, the Variational Quantum Eigensolver (VQE) with a TwoLocal ansatz and Variational Quantum Deflation (VQD), are used to compute ground and excited state energies. The proposed approach achieves high accuracies with an error of only 1.11\% compared to exact methods. Notably, VQE outperforms classical approximations such as perturbation theory (error 6.71\%) and the Wentzel-Kramers-Brillouin (WKB) approximation (error 5.36\%), yielding more precise energy values. These results highlight the potential of quantum simulation and VQE as effective tools for investigating complex quantum systems, paving the way for future applications in quantum chemistry and materials science as quantum hardware continues to advance.
\end{abstract}
\maketitle
\section{Introduction} 
Simulating quantum mechanical systems has long been recognized as one of the central goals of quantum computing and a fundamental challenge in modern science   \cite{georgescu2014quantum}. Feynman first highlighted the inefficiency of classical computers in modeling quantum phenomena, noting that \textit{``Nature isn't classical, dammit… and if you want to make a simulation of nature, you'd better make it quantum mechanical"}   \cite{hey2018feynman}. He proposed quantum computers as devices that naturally evolve according to quantum principles, thereby enabling efficient simulation of other quantum systems. Quantum simulation has since emerged as a transformative application of quantum computing, motivated by the exponential complexity of solving the Schrodinger equation on classical machines   \cite{nielsen2010quantum, njokitransformative}. Even the most advanced supercomputers face severe limitations when dealing with large Hilbert spaces, as the dimensionality grows exponentially with the number of particles or degrees of freedom   \cite{wiki:curseofdimensionality, mendoza_arenas_quantum_nodate}. Lloyd further established that quantum computers can efficiently simulate any local quantum system, laying the theoretical foundation for digital quantum simulation through quantum gates   \cite{lloyd1996universal}. Early implementations relied on analog quantum simulators such as nuclear magnetic resonance (NMR), trapped ions, and optical lattices to explore models like the Ising and Hubbard Hamiltonians   \cite{greiner2002quantum, blatt2008entangled}. A significant milestone was achieved when Aspuru-Guzik et al. demonstrated that quantum algorithms could estimate molecular ground-state energies more efficiently than classical approaches   \cite{aspuru2005simulated}. Since then, rapid advances in quantum hardware including superconducting qubits (transmon, Xmon, flux, and phase qubits) and trapped-ion platforms (Yb\textsuperscript{+}, Ca\textsuperscript{+}, Be\textsuperscript{+}, Sr\textsuperscript{+}) have propelled the field toward practical quantum simulation   \cite{preskill2018quantum}. The quantum harmonic oscillator (QHO) serves as a fundamental model in quantum physics, providing exact solutions and useful approximations near equilibrium   \cite{Balachandran2024}. However, real-world systems often exhibit anharmonicity, where higher-order interactions cause frequency shifts, nonlinear couplings, and phenomena such as, thermal expansion, and overtones in vibrational spectra   \cite{Mandal2021}. Accurately modeling anharmonic potentials is therefore essential for understanding chemical bonding, spectroscopy, and condensed matter systems.

Hybrid quantum-classical algorithms have emerged as powerful tools for quantum simulation, with the Variational Quantum Eigensolver (VQE) standing out as one of the most promising approaches   \cite{cerezo2021variational}. VQE leverages parameterized quantum circuits (ansatz) whose parameters are optimized by classical routines to approximate ground and excited state energies. First demonstrated on photonic quantum hardware to estimate the ground-state energy of molecular hydrogen   \cite{peruzzo2014variational}, VQE has since proven effective in tackling nonlinear Hamiltonians, particularly in regimes where classical approaches such as perturbation theory or semiclassical methods like the Wentzel-Kramers-Brillouin (WKB) approximation fail   \cite{o2016scalable}. Despite known challenges including barren plateaus, noise sensitivity, and high measurement overhead   \cite{mcclean2018barren, rubin2018fermionic, J.2024} VQE remains a leading candidate for achieving quantum advantage in the noisy intermediate-scale quantum (NISQ) era. In this work, these methods are applied to simulate the Quantum Anharmonic Oscillator (QAHO) on IBM's Quantum Experience using a 3-qubit system. The QAHO Hamiltonian is represented through a quartic potential discretized in position space, which is then encoded into qubit states. To study system dynamics, QFT-based quantum circuits are designed capable of capturing quantum revivals, and anharmonic effects. Furthermore, VQE is employed with a TwoLocal ansatz and Variational Quantum Deflation (VQD) to compute ground and excited state energies, achieving high accuracy compared with exact diagonalization.

A closely related effort by Jain et al.   \cite{jain2021quantum} demonstrated one of the first quantum simulations of a QHO using IBM's Quantum Experience. Their approach encoded the QHO Hamiltonian into qubits and employed filter-based quantum circuits to track state oscillations and revivals. They further showed how extending the system to multiple qubits enhanced resolution, thereby validating the feasibility of representing continuous-variable models on finite qubit devices. However, their work remained confined to the harmonic regime, where oscillations occur with constant frequency and no higher-order interactions are present. In contrast, this study advances this framework by extending the simulation to the anharmonic regime, where nonlinear interactions significantly influence system dynamics. By combining circuit-based time evolution with VQE and VQD, anharmonic phenomena such as frequency shifts are not only captured, but precise energy estimations beyond the reach of conventional classical methods are also achieved. This establishes a pathway for leveraging hybrid quantum-classical algorithms in modeling more realistic physical systems relevant to chemistry and materials science.

\subsection{Novelty and Contributions}

The novelty of this study lies in the integration of quantum dynamics simulation with variational energy estimation for anharmonic systems, addressing limitations of prior harmonic-only or semi-classical approaches. Our key contributions are:
\begin{itemize}
    \item A filter-based circuit is designed with Toffoli gates to simulate the time evolution of a QAHO, capturing effects such as revivals on a 3-qubit system.
    \item The framework is generalized to n-qubit systems, enabling improved spatial resolution and extending applicability to more complex quantum systems.
    \item VQE is employed combined with deflation techniques to compute both ground and excited state energies with a mean absolute percentage error (MAPE) of only 1.11\% compared to exact diagonalization, significantly outperforming classical methods such as perturbation theory (MAPE 6.71\%) and WKB approximation (MAPE 5.36\%).
\end{itemize}

The paper is organized as follows: Section  \ref{SecII} provides the theoretical background of the QAHO and its Hamiltonian formulation. Section  \ref{SecIII} outlines the methodology, including discretization, circuit design, and VQE implementation. Section  \ref{SecIV} presents the results, showcasing system dynamics and energy estimations compared with classical methods. Finally, Section  \ref{SecV} concludes the study, highlighting key findings, limitations, and future directions.

\section{Background}\label{SecII}

The general equation for a particle in a one-dimensional anharmonic potential is expressed as
\begin{equation}
\hat{H} = \frac{\hat{p}^2}{2m} + \frac{1}{2} m \omega^2 \hat{x}^2 + \lambda \hat{x}^4
\end{equation}
where \(m\), \(\omega\), and \(\lambda\) represent the mass, angular frequency, and anharmonicity parameter, respectively. 
Extending this to two dimensions, the anharmonic oscillator potential is defined as
\begin{equation}
    V(x, y) = \frac{1}{2} m\omega^2 (x^2 + y^2) + \lambda (x^4 + y^4).
\end{equation}
The corresponding Hamiltonian is written as
\begin{equation}
\hat{H} = -\frac{\hbar^2}{2m} \left( \frac{\partial^2}{\partial x^2} + \frac{\partial^2}{\partial y^2} \right) + \frac{1}{2} m\omega^2 (x^2 + y^2) + \lambda (x^4 + y^4).
\end{equation}
For simplicity, by setting \( \hbar = m = \omega = 1 \), the Hamiltonian reduces to
\begin{equation}
\hat{H} = -\frac{1}{2} \left( \frac{\partial^2}{\partial x^2} + \frac{\partial^2}{\partial y^2} \right) + \frac{1}{2} (x^2 + y^2) + \lambda (x^4 + y^4).
\end{equation}
In terms of kinetic and potential energies, the Hamiltonian in two-dimensional space is expressed as
\begin{equation}
\hat{H} = \frac{1}{2}(p_{x}^2+p_{y}^2) + \frac{1}{2} (x^2 + y^2) + \lambda (x^4 + y^4).
\end{equation}
Applying perturbation theory up to the first-order correction, the approximate energy eigenvalues for both spatial quantum numbers are obtained as
\begin{equation}
\begin{aligned}
E_{n_x,n_y} \approx &\ \hbar \omega (n_x + n_y + 1) + 6\lambda \left( \frac{\hbar}{2m\omega} \right)^2 \\
&\times \left[ n_x^2 + n_y^2 + n_x + n_y + 1 \right].
\end{aligned}
\end{equation}
\begin{figure}
    \centering
    \includegraphics[width=\linewidth]{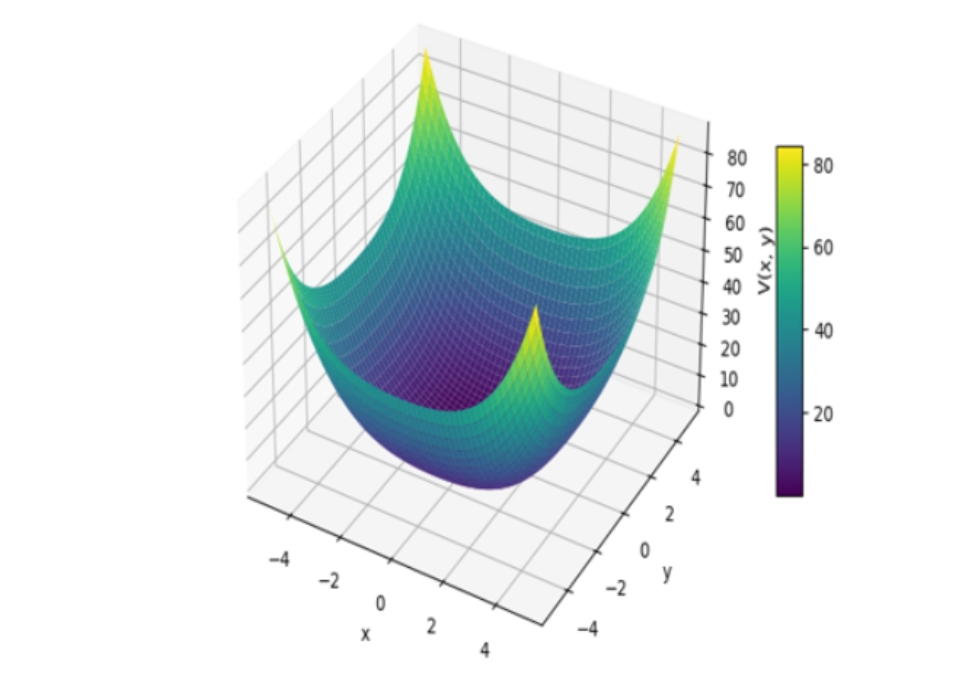}
    \caption{Potential surface of a 2D QAHO.}
    \label{fig:discretized_space}
\end{figure}
Discretization converts continuous functions and models into their discrete forms. In the simulation of a QAHO, discretization is necessary because quantum computers require computations with a finite number of elements in order to produce results within practical time frames. 
The discretized two-dimensional potential surface is shown in Fig.~\ref{fig:discretized_space}.
The discretized Hamiltonian of the system is given as
\begin{equation}
\hat{H}^d = \frac{1}{2}((\hat{p}_{x}^d)^2 + (\hat{p}_{y}^d)^2) + \frac{1}{2} (\hat{x}_d^2 + \hat{y}_d^2) + \lambda (\hat{x}_d^4 + \hat{y}_d^4),
\end{equation}
where \(\hat{p}_x^d\), \(\hat{p}_y^d\) denote the discrete momentum operators, and \(\hat{x}^d\), \(\hat{y}^d\) represent the discrete position operators for the \(x\) and \(y\) spatial coordinates. The discretized position operator for an \(N\)-dimensional Hilbert space   \cite{jain2021quantum} is defined as
\begin{eqnarray}
\hat{x}^d = \sqrt{\frac{\pi}{2N}} \cdot \text{diag}\left(-\frac{N}{2}, -\frac{N}{2}+1, \ldots, \frac{N}{2}-1\right).
\end{eqnarray}
The momentum operator \((\hat{p}_x^d)\) can be obtained by directly computing the momentum eigenvalues for each position state. 
A more effective method, however, employs the quantum Fourier transform (QFT), since the position operator is diagonal in the computational basis whereas the momentum operator becomes diagonal in the Fourier basis. 
The QFT transforms the wavefunction from position space to momentum space, where the momentum operator acts directly, and the inverse QFT returns it to position space. 
This procedure facilitates the construction of the discrete momentum operator, expressed as
\begin{equation}
    \hat{p}_d = (F_d)^{-1} \cdot \hat{x}_d \cdot (F_d),
\end{equation}
where \(F_d\) is the discrete QFT matrix, with elements defined as
\begin{equation}
    [F_d]_{j,k} = \frac{e^{\frac{2\pi ijk}{N}}}{\sqrt{N}}.
\end{equation}
This formulation is advantageous since efficient QFT circuits have been established for \(n\)-qubit systems. The time evolution of a quantum state is governed by the time translation operator
\begin{equation}
\hat{T} = e^{-\frac{i\hat{H}^dt}{\hbar}},
\end{equation}
where the Hamiltonian serves as the generator of time translation. 
Accordingly, the unitary operator describing the time evolution of position eigenstates is given by
\begin{equation}
U(t) = e^{-i\hat{H}^d t},
\end{equation}
which can also be expressed as
\begin{equation}
U(t) = e^{-it\left[\frac{1}{2}(p_{x}^2 + p_{y}^2) + \frac{1}{2}(x^2 + y^2) + \lambda (x^4 + y^4)\right]}.
\end{equation}
Restricting the operator to its \(x\)-dependence yields
\begin{equation}
U_x(t) = e^{\left[ \frac{-it}{2} \left( ((F^d)^{-1} \cdot [x^d]^2 \cdot F^d) + ([x^d]^2 + 2\lambda[x^d]^4) \right) \right]}.
\end{equation}
The position operator \([x^d]\) is therefore the only matrix required to evaluate \(U(t)\). 
Since \([x^d]\) is diagonal, it can be expanded using the matrix exponential as
\begin{equation}
    e^{\left( -\frac{ut}{2} [A] \right)} = \mathbb{I} + \sum_{m=1}^{\infty} \left( -\frac{ut}{2} \right)^m \frac{[A]^m}{m!},
\end{equation}
where \(A\) denotes the corresponding operator matrix. As \([x^d]\) contains a finite number of elements, the expansion is convergent, enabling time evolution for any arbitrary state. Such an evolution can be realized using \(n\)-qubit quantum circuits, allowing the simulation to be generalized to higher-dimensional systems. Unitary operators govern quantum system dynamics, ensure probability conservation, and permit the embedding of non-unitary processes within extended Hilbert spaces. From Eq 1 , the Hamiltonian can also be written as
\begin{eqnarray}
H = \frac{p^2}{2m} + \frac{1}{2} m \omega^2 x^2 + \lambda x^4.
\end{eqnarray}
Introducing the ladder operators \(a, a^\dagger\), the position and momentum operators take the forms
\begin{equation}
x = \sqrt{\frac{\hbar}{2m\omega}}(a + a^\dagger), \quad p = i \sqrt{\frac{\hbar m \omega}{2}} (a^\dagger - a).
\end{equation}
For \(\hbar = 1, m = 1, \omega = 1\), the Hamiltonian simplifies to
\begin{equation}
H = a^\dagger a + \frac{1}{2} + \lambda \left( \frac{1}{\sqrt{2}} (a + a^\dagger) \right)^4,
\end{equation}
which reduces further to
\begin{equation}
H = a^\dagger a + \frac{1}{2} + \lambda \cdot \frac{1}{4}(a + a^\dagger)^4.
\end{equation}

The Hilbert space may be truncated to eight energy levels using a 3-qubit system. 
These states form the basis \(\{ \ket{0}, \ket{1}, \ldots, \ket{7} \}\), which can be mapped onto the computational basis of three qubits as
\begin{equation}
\begin{aligned}
\ket{000} &\equiv \ket{0}, &\quad \ket{001} &\equiv \ket{1}, \\
\ket{010} &\equiv \ket{2}, &\quad \ket{011} &\equiv \ket{3}, \\
\ket{100} &\equiv \ket{4}, &\quad \ket{101} &\equiv \ket{5}, \\
\ket{110} &\equiv \ket{6}, &\quad \ket{111} &\equiv \ket{7}.
\end{aligned}
\end{equation}

In this truncated 8-dimensional Fock basis, the annihilation operator \(a\) is represented as
\begin{equation}
a = \begin{bmatrix}
0 & \sqrt{1} & 0 & 0 & 0 & 0 & 0 & 0 \\
0 & 0 & \sqrt{2} & 0 & 0 & 0 & 0 & 0 \\
0 & 0 & 0 & \sqrt{3} & 0 & 0 & 0 & 0 \\
0 & 0 & 0 & 0 & \sqrt{4} & 0 & 0 & 0 \\
0 & 0 & 0 & 0 & 0 & \sqrt{5} & 0 & 0 \\
0 & 0 & 0 & 0 & 0 & 0 & \sqrt{6} & 0 \\
0 & 0 & 0 & 0 & 0 & 0 & 0 & \sqrt{7} \\
0 & 0 & 0 & 0 & 0 & 0 & 0 & 0
\end{bmatrix}.
\end{equation}
The Hamiltonian is then given by
\begin{equation}
H = a^\dagger a + \frac{1}{2} + \lambda \cdot \frac{1}{4}(a + a^\dagger)^4,
\end{equation}
and substituting \(\lambda = 0.05\) yields
\begin{equation}
H = a^\dagger a + \frac{1}{2} + \frac{0.05}{4}(a + a^\dagger)^4.
\end{equation}
The resulting Hamiltonian matrix takes the form
\begin{equation}
\resizebox{\columnwidth}{!}{$
H = \begin{bmatrix}
0.5375 & 0 & 0.1061 & 0 & 0.0612 & 0 & 0 & 0 \\
0 & 1.6875 & 0 & 0.3062 & 0 & 0.1369 & 0 & 0 \\
0.1061 & 0 & 2.9875 & 0 & 0.6062 & 0 & 0.2372 & 0 \\
0 & 0.3062 & 0 & 4.4375 & 0 & 1.0062 & 0 & 0.3623 \\
0.0612 & 0 & 0.6062 & 0 & 6.0375 & 0 & 1.5062 & 0 \\
0 & 0.1369 & 0 & 1.0062 & 0 & 7.7875 & 0 & 1.4582 \\
0 & 0 & 0.2372 & 0 & 1.5062 & 0 & 8.9875 & 0 \\
0 & 0 & 0 & 0.3623 & 0 & 1.4582 & 0 & 8.6375 \\
\end{bmatrix}
$}.
\end{equation}

\section{Methodology}\label{SecIII} 
\subsection{Quantum Simulation}

A 3-qubit system corresponds to a Hilbert space of dimension \(2^3 = 8\), which in two spatial dimensions (\(x, y \in [-8, 8]\)) yields \(8 \times 8 = 64\) mesh points. The discretized position operator \(\hat{x}^d\) is therefore given as:
\begin{eqnarray}
[x^d] &=& 
\begin{bmatrix}
-4 & 0 & \cdots & 0 \\
0 & -3 & \ddots & \vdots \\
\vdots & \ddots & 2 & 0 \\
0 & \cdots & 0 & 3
\end{bmatrix}, \
{[x^d]}^2 = 
\begin{bmatrix}
16 & 0 & \cdots & 0 \\
0 & 9 & \ddots & \vdots \\
\vdots & \ddots & 4 & 0 \\
0 & \cdots & 0 & 9
\end{bmatrix},\nonumber\\
{[x^d]}^4 &=& 
\begin{bmatrix}
256 & 0 & \cdots & 0 \\
0 & 81 & \ddots & \vdots \\
\vdots & \ddots & 16 & 0 \\
0 & \cdots & 0 & 81
\end{bmatrix}.
\end{eqnarray}
\begin{table}[h]
\centering
\caption{Diagonal entries of the time evolution operator $U_{\hat{x}}(t)$.}
\label{tab:time_evo}
\begin{tabular}{|c|l|} % Changed to l to avoid parsing issues; use manual wrapping if needed
\hline
\textbf{Entry} & \textbf{Series Expansion and Approximate Exponential} \\
\hline
$U_{\hat{x}}(t)[1,1]$ & $1 - 2.0643it - 2.1315t^2 + \cdots = e^{-2.064it}$ \\
\hline
$U_{\hat{x}}(t)[2,2]$ & $1 - 1.0398it - 0.5406t^2 + \cdots = e^{-1.039it}$ \\
\hline
$U_{\hat{x}}(t)[3,3]$ & $1 - 0.4235it - 0.0897t^2 + \cdots = e^{-0.423it}$ \\
\hline
$U_{\hat{x}}(t)[4,4]$ & $1 - 0.1001it - 0.0050t^2 + \cdots = e^{-0.100it}$ \\
\hline
$U_{\hat{x}}(t)[5,5]$ & $1$ \\
\hline
$U_{\hat{x}}(t)[6,6]$ & $1 - 0.1001it - 0.0050t^2 + \cdots = e^{-0.100it}$ \\
\hline
$U_{\hat{x}}(t)[7,7]$ & $1 - 0.4235it - 0.0897t^2 + \cdots = e^{-0.423it}$ \\
\hline
$U_{\hat{x}}(t)[8,8]$ & $1 - 1.0398it - 0.5406t^2 + \cdots = e^{-1.039it}$ \\
\hline
\end{tabular}
\end{table}

\begin{figure}[h]
\centering
\begin{tabular}{cc}
(a) & (b) \\
\includegraphics[width=0.45\linewidth]{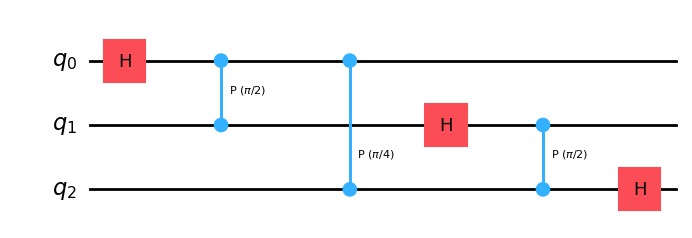} &
\includegraphics[width=0.45\linewidth]{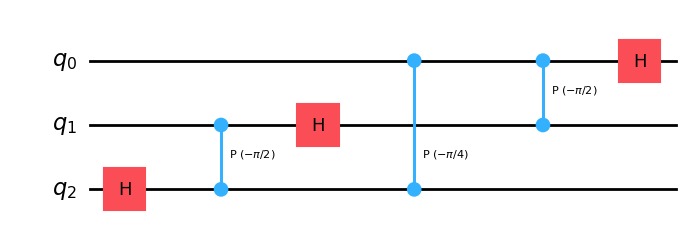} \\
\end{tabular}
\caption{Quantum circuit depicting (a) QFT and (b) IQFT.}
\label{fig:qft_iqft}
\end{figure}

The corresponding time evolution operator is expressed as:
\begin{equation}
U_x(t) = e^{-\frac{i t}{2} \left( \hat{F}^{\dagger} (x^d)^2 \hat{F} \right)} \cdot e^{-\frac{i t}{2} \left((x^d)^2 + 2\lambda(x^d)^4 \right)} \label{time_evo_x}
\end{equation}
For small anharmonicity (\(\lambda = 0.05\)), the unitary operator governing kinetic and potential energy contributions can be evaluated. Applying the operator to the potential energy part yields the diagonal entries shown in Table~\ref{tab:time_evo}. From Eq.~\eqref{time_evo_x}, the operator takes the form
\begin{eqnarray}
U_{\hat{x}}(t) &=& 
\begin{bmatrix}
e^{-2.06i t} & 0 & \cdots & 0 \\
0 & e^{-1.039i t} & \ddots & \vdots \\
\vdots & \ddots & e^{-0.423i t} & 0 \\
0 & \cdots & 0 & e^{-1.039i t}
\end{bmatrix}, \nonumber \\
&=& e^{-2.064i t}
\begin{bmatrix}
1 & 0 & \cdots & 0 \\
0 & e^{-1.024i t} & \ddots & \vdots \\
\vdots & \ddots & e^{-1.640i t} & 0 \\
0 & \cdots & 0 & e^{-1.024i t}
\end{bmatrix}.
\end{eqnarray}

\begin{figure}
\centering
\includegraphics[width=0.5\textwidth]{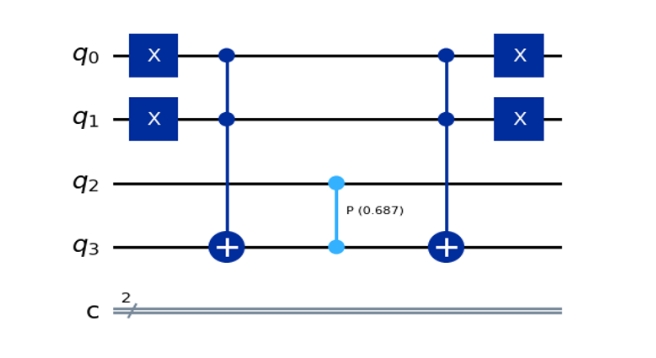}
\caption{Filter circuit for state \(\ket{001}\).}
\label{fig:filter-001}
\end{figure}
\begin{figure*}
\centering
\includegraphics[width=1.\linewidth]{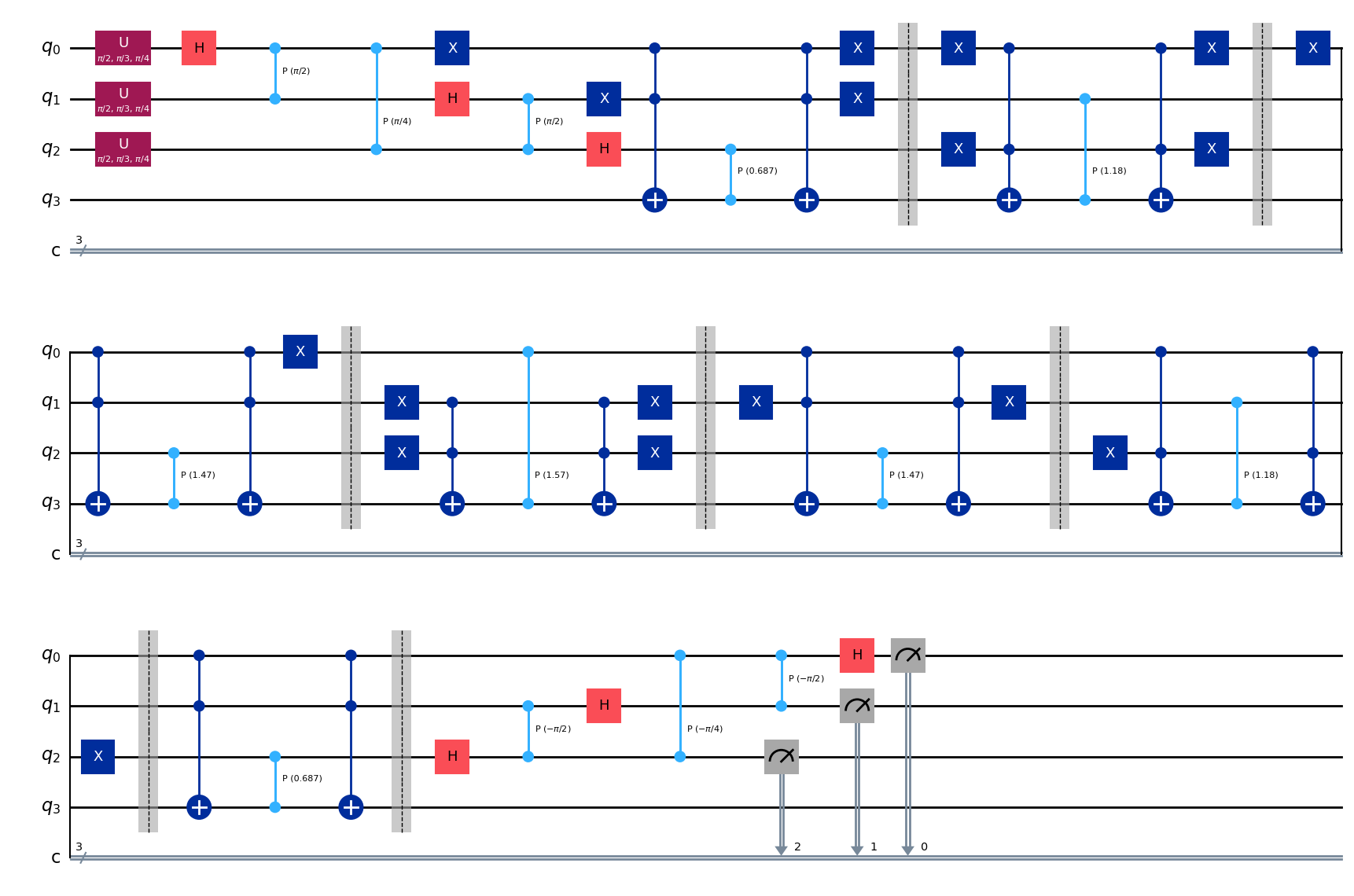}
\caption{Quantum circuit for \(U_{\hat{p}}(t)\) implementation via QFT sandwiching and filter-based design.}
\label{fig:kinetic-circuit}
\end{figure*}

The first matrix element is \(1\), meaning the \(\ket{000}\) state remains unchanged, thereby avoiding undesired phases. The other basis states acquire phases determined by their position values, enabling accurate modeling of physical phenomena. The remaining diagonal entries provide the corresponding phase factors for the other seven basis states. The unitary operator for kinetic energy simulation in a 3-qubit system is implemented using QFT and inverse QFT circuits, as illustrated in Fig.~\ref{fig:qft_iqft}. This approach forms the basis for simulating time evolution under Hamiltonians dependent on both position and momentum. The complete unitary time-evolution operator is expressed as
\begin{equation}
U(t) = U_{p_x} \cdot U_{p_y} \cdot U_{x\text{anh}} \cdot U_{y\text{anh}},
\end{equation}
which makes the implementation modular and computationally efficient. This method provides a systematic approach to constructing and applying unitary operators for arbitrary quantum states in multi-dimensional systems, laying the foundation for quantum circuit-based simulation of quantum dynamics. The quantum circuit is constructed using QFT, inverse QFT, and a sequence of Toffoli gates. The single-qubit controlled-rotation gate is represented by the matrix
\begin{equation}
CU1_n = \begin{bmatrix}
1 & 0 \\
0 & e^{2\pi i / 2^n}
\end{bmatrix}
\end{equation}
with the corresponding phase angle
\begin{equation}
\theta_n = \frac{2\pi}{2^n}.
\end{equation}

For a 3-qubit system, the position operator \(x^d\) is diagonal with \(2^3 = 8\) eigenvalues, mapped directly onto the diagonal of the unitary operator \(U_{x^d} = e^{-i x^d t}\), where each diagonal entry corresponds to a phase factor. The circuit implementing this operator employs a filter-based architecture in which qubits are selectively addressed through controlled and anti-controlled gates. The third qubit’s state is transferred to an ancilla, enabling a controlled phase rotation on another qubit. A mirrored sequence of Toffoli gates resets intermediate states, ensuring the next filter acts on a clean input. Initialization in a computational basis state localizes the particle’s position, and its time evolution is observed through the probability distribution across basis states. The filter circuit for state \(\ket{001}\) is presented in Fig.~\ref{fig:filter-001}, while the full circuit for \(U_{\hat{p}}(t)\) (kinetic energy simulation) is shown in Fig.~\ref{fig:kinetic-circuit}. This architecture generalizes naturally to an \(n\)-qubit system by extending the number of filters to \((n-1)\). The operator \(U_{\hat{x}}^{[n]}\) is implemented using \((n-1)\) such filters, whereas \(U_{\hat{p}}^{[n]}\) requires the insertion of an additional \((n-1)\) filters between the QFT and its inverse. All simulations are performed in {Python} using the {Qiskit} framework on the {QASM simulator} backend, with 8192 shots to ensure statistical accuracy.

\subsection{Variational Quantum Eigensolver (VQE)}
VQE is a hybrid algorithm designed to estimate the ground and excited state energies of quantum systems. A quantum processor is employed to prepare and evaluate quantum states, while a classical computer optimizes the circuit parameters. This division of tasks enables VQE to remain effective even on current noisy hardware with limited coherence times    \cite{mcclean2016theory}. By keeping quantum circuits shallow and delegating optimization to classical routines, VQE achieves reliable performance despite hardware imperfections    \cite{kandala2017hardware}. Its hybrid nature, combining quantum state preparation with classical optimization, makes it particularly well-suited for the noisy intermediate-scale quantum (NISQ) era. In quantum computation, Hamiltonians are expressed in the Pauli basis, i.e., as linear combinations of Pauli strings such as IXZ, XYY, etc. This representation facilitates efficient simulation on quantum devices. Qiskit provides a compact format called \texttt{SparsePauliOp}, which improves computational efficiency for quantum circuit evaluation and expectation value calculations. An ansatz is a parameterized quantum circuit designed to approximate eigenstates of the Hamiltonian. The \texttt{TwoLocal} ansatz incorporates alternating layers of single-qubit rotations (e.g., \(R_y\), \(R_z\)) and entangling gates (e.g., CZ), offering a flexible yet hardware-efficient architecture that balances expressivity and implementation feasibility. The estimator is used to compute the expectation value of the parameterized circuit.

\begin{equation}
\langle \psi(\theta) | H | \psi(\theta) \rangle
\end{equation}

To ensure orthogonality when extracting excited states, fidelity calculations are performed using the Compute-Uncompute method. This step is essential in the deflation process employed by VQD. VQD extends VQE by introducing orthogonality constraints that allow systematic determination of excited states. The algorithm operates iteratively: it first identifies the lowest-energy state, then augments the cost function with a penalty term for overlap with previously computed states, and finally minimizes the updated objective. This iterative procedure ensures accurate extraction of both ground and excited states.

\subsection{Perturbation theory}
From Eq. (22), the Hamiltonian is obtained as  
\begin{equation}
V(x) = \frac{1}{2}x^2 + \lambda x^4
\end{equation}
and the full Hamiltonian is expressed as  
\begin{equation}
\hat{H} = -\frac{1}{2} \frac{d^2}{dx^2} + \frac{1}{2}x^2 + \lambda x^4.
\end{equation}
This form is written in natural units with \(\hbar = 1\) and \(m = 1\), such that both time and space are dimensionless. The potential describes a harmonic oscillator perturbed by a quartic term, where \(\lambda\) is a dimensionless coupling constant governing the strength of the anharmonicity. The unperturbed system corresponds to the quantum harmonic oscillator, with eigenstates and energies given by  
\begin{eqnarray}
E_n^{(0)} &=& \left( n + \frac{1}{2} \right),\nonumber\\
\psi_n(x) &=& \frac{1}{\sqrt{2^n n!}} \left( \frac{1}{\pi^{1/2}} \right)^{1/2} e^{-x^2/2} H_n(x).
\end{eqnarray}

Applying first-order perturbation theory, the energy correction due to the \(\lambda x^4\) term is expressed as  
\begin{equation}
E_n^{(1)} = \lambda \langle n | x^4 | n \rangle,
\end{equation}
where the expectation value is known analytically:  
\begin{equation}
\langle n | x^4 | n \rangle = \frac{3}{4}(2n^2 + 2n + 1).
\end{equation}

Thus, the first-order corrected energy becomes  
\begin{equation}
E_n \approx \left( n + \frac{1}{2} \right) + \lambda \cdot \frac{3}{4}(2n^2 + 2n + 1),
\end{equation}
and the general expression for the total energy is  
\begin{equation}
E_n(\lambda) = \left( n + \frac{1}{2} \right) + \lambda \cdot \frac{3}{4}(2n^2 + 2n + 1).
\end{equation}
For \(\lambda = 0.05\), the energies corresponding to \(n = 0\) through \(n = 7\) are obtained from  
\begin{equation}
E_n(0.05) = \left( n + \frac{1}{2} \right) + 0.05 \cdot \frac{3}{4}(2n^2 + 2n + 1).
\end{equation}

\subsection{WKB Approximation Method}

Consider the potential  
\begin{equation}
V(x) = \frac{1}{2}x^2 + \lambda x^4
\end{equation}
The WKB quantization condition provides approximate energy levels through  
\begin{equation}
\int_{x_1}^{x_2} \sqrt{2(E - V(x))} \, dx = \left(n + \frac{1}{2} \right)\pi,
\end{equation}
where \(x_1\) and \(x_2\) denote the classical turning points. For the quartic anharmonic oscillator, an approximate closed-form expression for the energy is given by  
\begin{equation}
E_n^{\text{WKB}} \approx \left(n + \frac{1}{2} \right) + \frac{3\lambda}{2} \left(n + \frac{1}{2} \right)^2.
\end{equation}
For \(\lambda = 0.05\), this becomes  
\begin{equation}
E_n^{\text{WKB}} = \left(n + \frac{1}{2} \right) \left[1 + \frac{3}{2} \cdot 0.05 \cdot \left(n + \frac{1}{2} \right)\right].
\end{equation}

\section{Results}\label{SecIV}

\begin{figure*}
    \centering
    \includegraphics[width=0.65\linewidth]{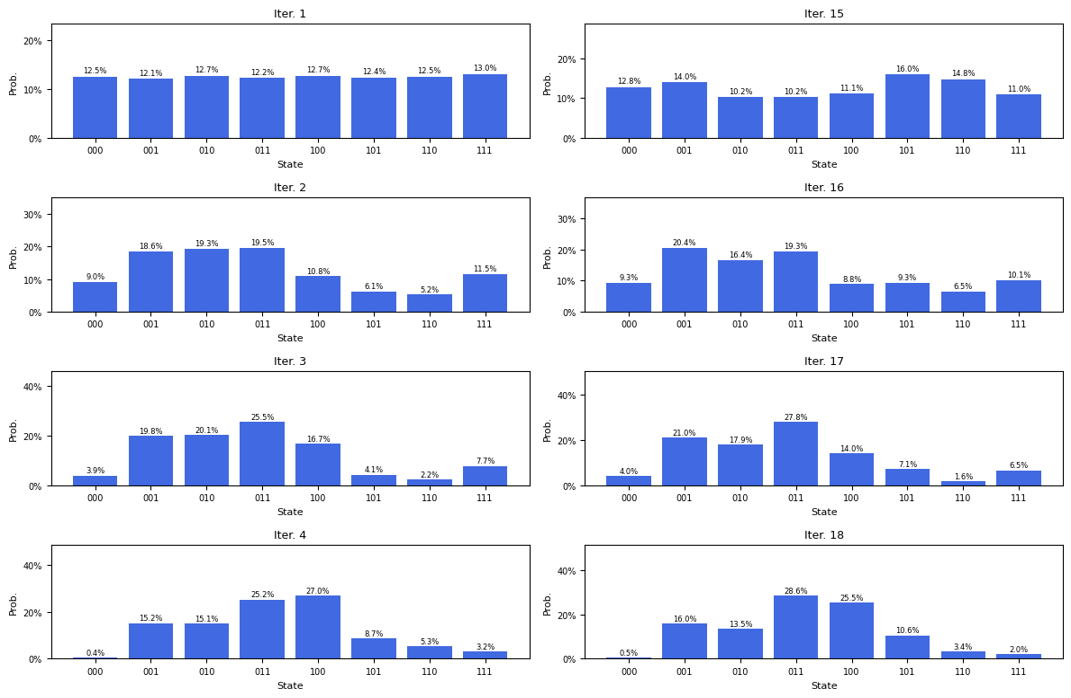}
    %\caption{Results after 18 iterations}
    \label{fig:fig1b}
    \includegraphics[width=0.65\linewidth]{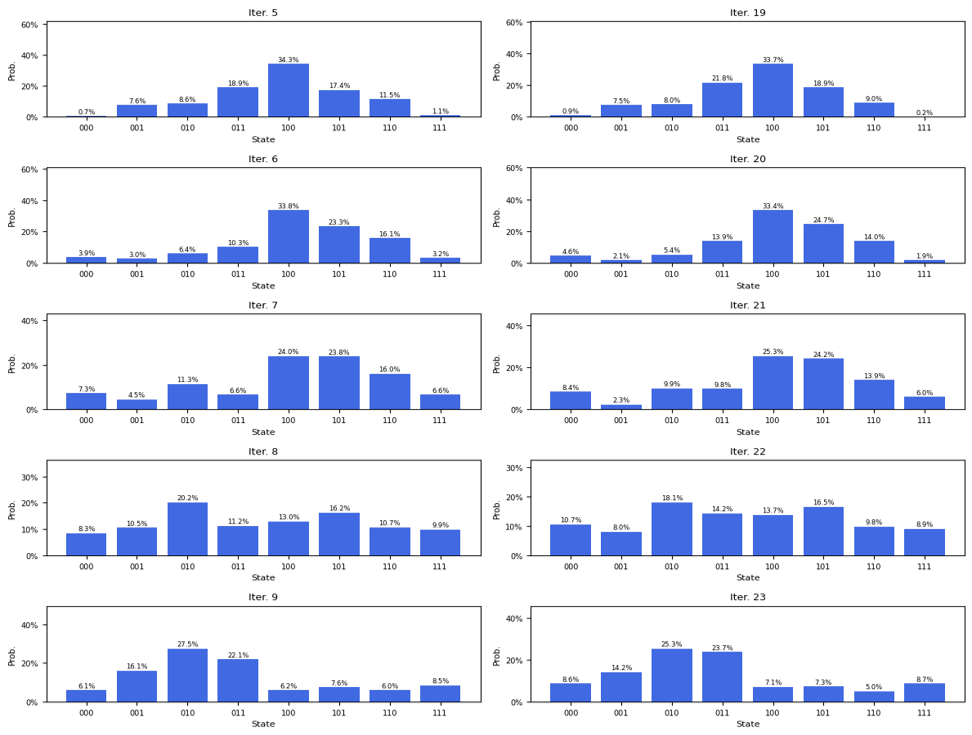}
    %\caption{Results after 23 iterations}
    \label{fig:fig2}
    \includegraphics[width=0.65\linewidth]{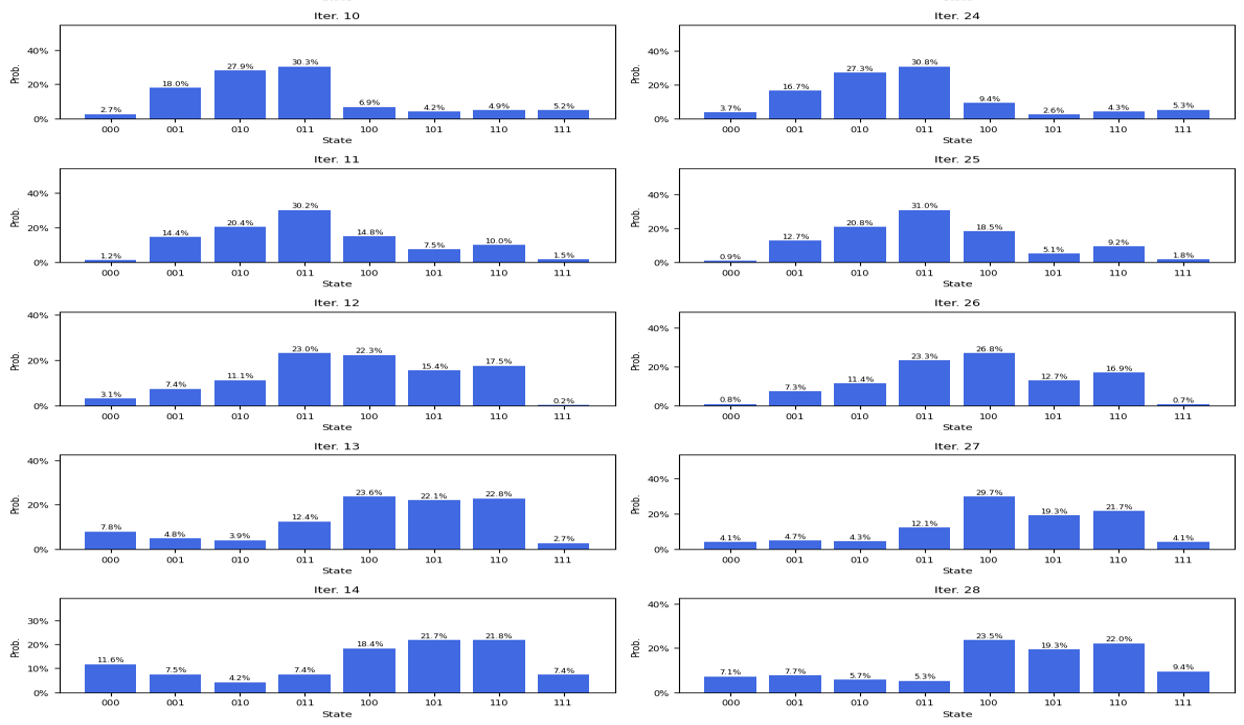}
    \caption{Probability amplitude from $0$ to $6\pi$.}
    \label{fig:fig3}
\end{figure*}
\begin{figure}[]
    \centering
    \includegraphics[width=\linewidth]{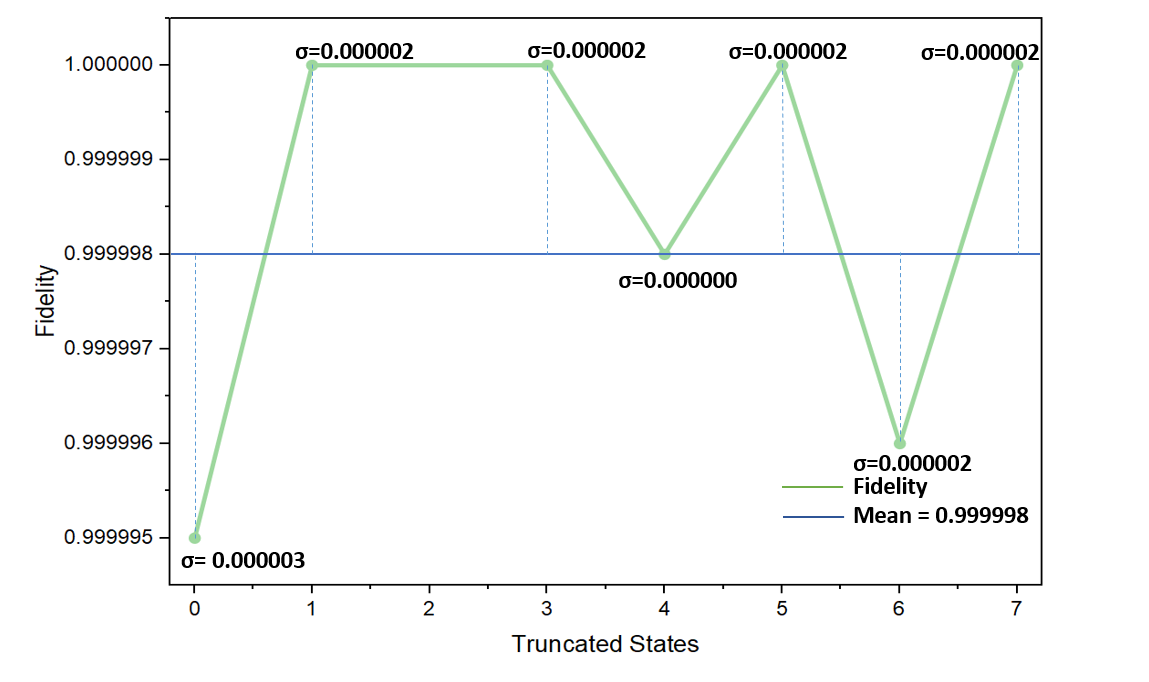}
    \caption{Fidelity curve for VQE method.}
    \label{fig:fidelity}
\end{figure}
\begin{figure*}[]
    \centering
    \includegraphics[width=\linewidth]{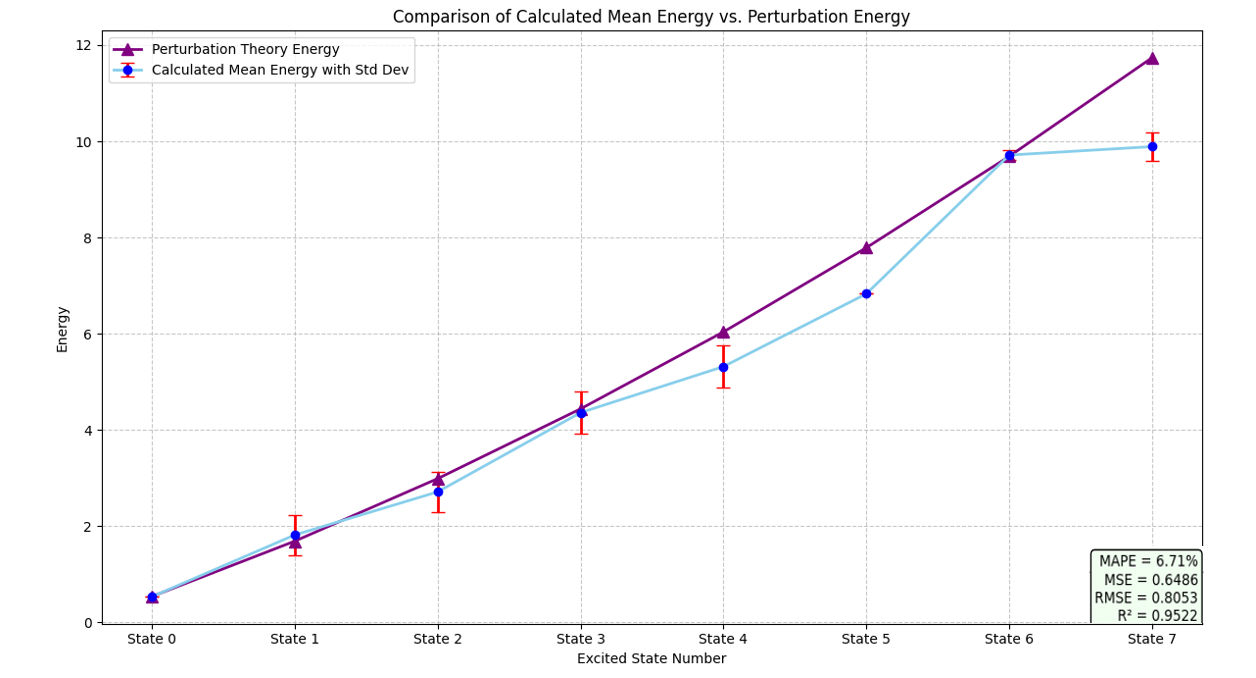}
    \caption{Comparison between VQE and perturbation method. Error metrics are shown in the annotation.}
    \label{fig:vqe-perturb}
\end{figure*}
\begin{figure*}[]
    \centering
    \includegraphics[width=\linewidth]{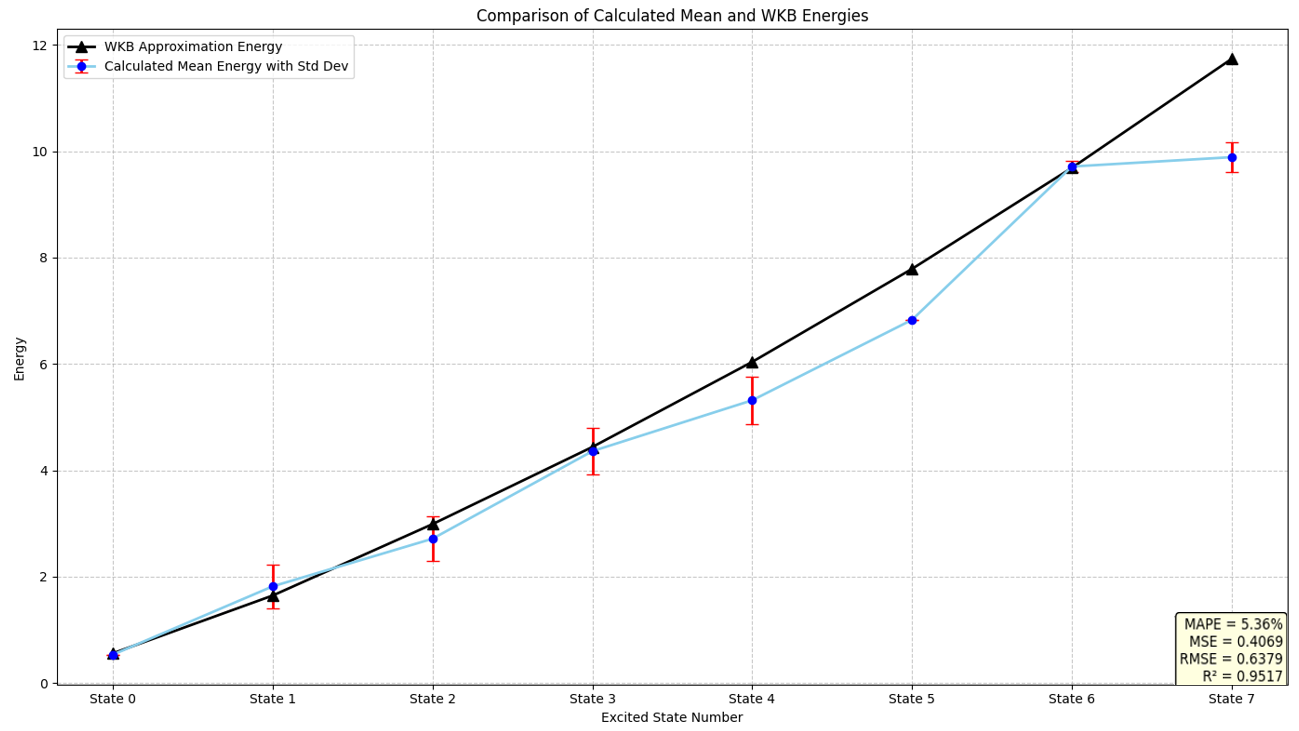}
    \caption{Comparison between VQE and WKB approximation. Error margins are visualized.}
    \label{fig:vqe-wkb}
\end{figure*}
\begin{figure*}
    \centering
    \includegraphics[width=\linewidth]{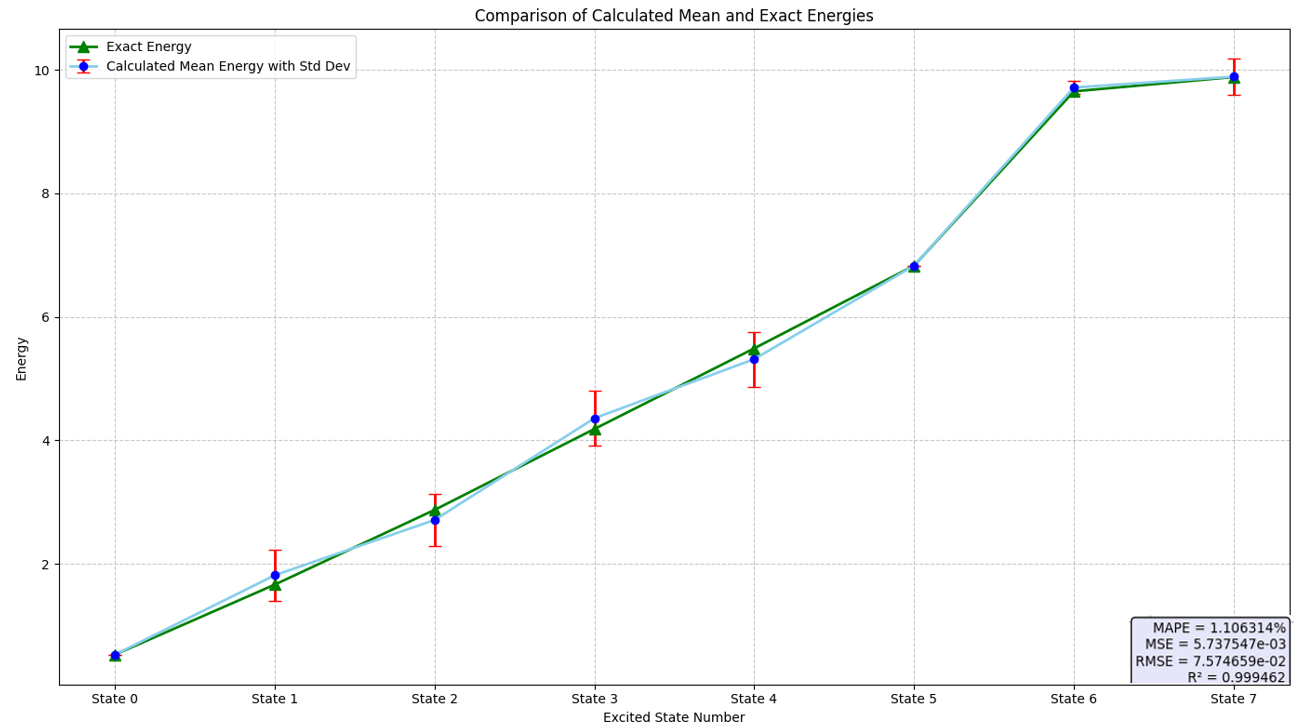} 
    \caption{Comparison between VQE and Exact diagonalization. The energies show strong agreement.}
    \label{fig:vqe-exact-fig}
\end{figure*} 

The simulation tracks a 3-qubit system mimicking an anharmonic oscillator over 28 steps, covering a time range of $t = 0 - 6\pi$, as illustrated in Fig.~\ref{fig:fig3}. It examines how the system evolves and highlights quantum revivals arising from interference. In Phase I (steps 1-14), the system begins in a low-energy configuration, predominantly occupying states such as $\ket{000}$, $\ket{001}$, $\ket{010}$, $\ket{011}$, and $\ket{100}$, indicating a concentrated probability distribution. From $t \approx \frac{\pi}{6}$ to $\frac{5\pi}{6}$, higher-energy states including $\ket{101}$ and $\ket{110}$ begin to acquire probability, suggesting the onset of spreading. By steps 6-10 ($t \approx \pi$ to $2\pi$), the anharmonic effect induces stronger mixing, with significant increases in the populations of $\ket{110}$ and $\ket{111}$, leading to dephasing caused by uneven energy levels. During steps 11-14, the probabilities across all states become nearly uniform, reflecting maximum spreading and a balanced distribution. In Phase II (steps 15-28), quantum revivals emerge. Between steps 15-18 ($t \approx 3\pi$ to $4\pi$), low-energy states such as $\ket{010}$, $\ket{011}$, and $\ket{100}$ regain amplitude, demonstrating partial recovery of the initial state through interference. In steps 19-24, probabilities redistribute toward higher-energy states once again due to phase mismatches. Finally, during steps 25-28 ($t \approx 5\pi$ to $6\pi$), the amplitudes of low-energy states increase once more, marking a second revival. The presence of the anharmonic potential ($\lambda = 0.05$) produces uneven energy gaps, resulting in dephasing and imperfect revivals. Nevertheless, quantum coherence enables the emergence of cyclic patterns. These results demonstrate how even a small 3-qubit system can capture complex quantum behaviors such as spreading, dephasing, and revivals, providing valuable insights into the study of real quantum systems. While simulating the ansatz through a classical optimizer, we obtain the average energy eigenvalues corresponding to all possible truncated states of the 3-qubit system. The fidelity values for the different states ($n=0$ through $n=7$) remain consistently high, clustering tightly around a mean fidelity of $0.999999$ (Fig. \ref{fig:fidelity}). This near-perfect average demonstrates excellent agreement between the VQE results and exact diagonalization, thereby confirming the accuracy of VQE in modeling the quantum states of the anharmonic oscillator. The fidelity of most individual states lies extremely close to the mean value, which shows that the VQE method reliably reproduces each quantum state with only minimal deviations. Furthermore, all state fidelities fall within the $\pm 1$ standard deviation range, indicating that fluctuations are negligible and that the performance is stable across the entire spectrum of states. The narrow spread of fidelities highlights the robustness of the VQE approach in matching exact solutions, underlining its reliable accuracy for simulating both ground and excited states in complex quantum systems. In addition, from Eq. (39) we derive the table of energy eigenvalues for the different truncated states, while Eq.~(43) provides another corresponding table of energy eigenvalues, further validating the consistency of the obtained results across different formulations.

We now compare the energy eigenvalues obtained from the VQE with those from perturbation theory, the WKB approximation, and exact diagonalization. The results of this comparison are summarized in Table~\ref{tab:comparison_methods} and illustrated in the corresponding figures. The plot shown in Fig.~\ref{fig:vqe-perturb} presents a direct comparison of VQE and perturbation theory across quantum numbers $n=0$ through $n=7$. The VQE energies (blue line) consistently exceed the perturbation theory estimates (purple dashed line with crosses), with the deviation growing at higher quantum numbers and reaching nearly one energy unit at $n=7$. Despite this, both methods follow a similar overall trend, increasing monotonically from values near zero at $n=0$ to approximately $10$ (VQE) and $11$ (perturbation) at $n=7$. Quantitative error metrics provide additional insight into the degree of agreement. The MAPE is $6.71\%$, which reflects a moderate relative discrepancy between the two methods. The mean squared error (MSE) of $0.6486$ and root mean squared error (RMSE) of $0.8053$ indicate an average absolute deviation of roughly $0.805$ energy units on a 0-12 scale. Furthermore, the coefficient of determination, $R^2 = 0.6322$, shows that VQE explains approximately $63\%$ of the variance in perturbation theory energies, pointing to a moderate correlation. These results reveal a systematic tendency of VQE to overestimate the energy eigenvalues when compared with perturbation theory. The fact that the deviation grows with higher quantum states suggests that scalability challenges emerge, possibly due to limitations in the chosen ansatz or classical optimizer. Although the MAPE and RMSE values indicate that VQE reproduces the general trend, the observed pattern of overestimation emphasizes the need for methodological refinements to improve accuracy and achieve closer agreement with perturbative reference results.

The graph shown in Fig.~\ref{fig:vqe-wkb} illustrates a comparison of the energy values obtained from the VQE and the WKB approximation for truncated states $n=0$ through $n=7$. The VQE energies (blue line) are consistently higher than the WKB energies (green dashed line), and the deviation becomes more pronounced as the quantum number increases. By the seventh state, the difference reaches nearly 1.5 energy units. Despite this discrepancy, both methods display the same overall monotonic trend, rising from near zero at $n=0$ to approximately 10 for VQE and 8.5 for WKB at $n=7$. Quantitative error analysis provides further insight. The MAPE is $5.36\%$, indicating a relatively small relative discrepancy between the two methods. The MSE is $0.4069$, and the RMSE is $0.6379$, showing a moderate absolute deviation of about $0.64$ energy units. The coefficient of determination, $R^2 = 0.9517$, demonstrates that the VQE results strongly correlate with the WKB estimates, capturing approximately 95\% of the observed variance. Although the relative error is low, the increasing absolute deviations at higher quantum states reveal that VQE tends to systematically overestimate energies compared with WKB. This tendency may arise from limitations in the choice of variational ansatz or in the optimization strategy employed. Incorporating more expressive variational circuits or applying simple post-processing correction schemes could mitigate these discrepancies. Such refinements would further enhance VQE's reliability, positioning it as a more accurate and effective approach for estimating quantum energy levels compared to semiclassical approximations like WKB.

The graph shown in Fig.~\ref{fig:vqe-exact-fig} presents a comparison of energy estimates obtained from VQE and exact diagonalization across truncated states $n=0$ through $n=7$. The VQE energies (blue line) are consistently slightly higher than the exact energies (red dashed line), with the difference gradually increasing at higher quantum numbers and reaching a maximum of about 0.5 energy units at $n=7$. Both methods exhibit a similar rising trend, progressing from values near zero at $n=0$ to approximately 10 at $n=7$. The error metrics further quantify this agreement. The MAPE is $1.1063\%$, reflecting an extremely small relative discrepancy. The MSE of $5.7374 \times 10^{-3}$ and the RMSE of $7.5747 \times 10^{-2}$ correspond to an absolute deviation of only $0.076$ energy units, which is negligible on the 0-10 scale. Moreover, the coefficient of determination, $R^2 = 0.9999462$, indicates an almost perfect correlation, with VQE accounting for 99.99\% of the variability in the exact results. The standard deviation of residuals, $\sigma = 0.071840$, further confirms the tight fit between the two datasets. The small deviations observed, even at higher states, highlight VQE’s high degree of accuracy and its scalability for this system. The low MAPE and RMSE values demonstrate that VQE serves as a reliable alternative to exact diagonalization, closely approximating the true energy values with minimal error. For quantum energy estimation tasks, VQE performs exceptionally well and requires little to no refinement to achieve precision comparable to exact methods.

Table~\ref{tab:comparison_methods} presents a quantitative comparison of the energy eigenvalues obtained via different methods for the quantum anharmonic oscillator. Using exact diagonalization as the reference, the VQE results show the closest agreement, with deviations typically below $0.1$ energy units, for example $|\Delta E|\approx 0.011$ at the first excited state ($n=1$) and $|\Delta E|\approx 0.082$ at the second excited state ($n=2$). The largest VQE discrepancy occurs at $n=4$ with an error of about $0.162$, while for states $n=0$ and $n=5$ the agreement is essentially exact. By contrast, perturbation theory and WKB approximations systematically overestimate the energies at higher excitations: for instance, at $n=5$ both predict $\sim 7.79$ compared to the exact $6.83$, leading to an error close to $0.96$. At $n=7$, the overestimation reaches nearly $1.85$ for both methods. Nonetheless, perturbation and WKB are reasonably accurate for low-lying states ($n=0$-$2$) where errors remain $\lesssim 0.11$. Overall, the VQE method most faithfully reproduces the exact spectrum across the considered states, whereas perturbative and semiclassical approximations diverge significantly at higher excitations.

\begin{table}[htbp!]
\centering
\renewcommand{\arraystretch}{1.2}
\caption{Comparison of energy eigenvalues of the quantum anharmonic oscillator obtained using different methods: VQE    \cite{Higgott2019variationalquantum}, perturbation theory    \cite{Adelakun2014}, WKB approximation    \cite{Bhadra_2024}, and exact Diagonalization.}
\label{tab:comparison_methods}
\begin{tabular}{|c|c|c|c|c|}
\hline
\textbf{State} & \textbf{VQE} & \textbf{Perturbation} & \textbf{WKB} & \textbf{Exact} \\
\hline
0 & 0.532151   & 0.5375   & 0.557   & 0.53215009 \\
1 & 1.653929   & 1.6875   & 1.644   & 1.665392897 \\
2 & 2.794014   & 2.9875   & 2.987   & 2.87513953 \\
3 & 4.268678   & 4.4375   & 4.438   & 4.18508178 \\
4 & 5.329061   & 6.0375   & 6.038   & 5.49052597 \\
5 & 6.829527   & 7.7875   & 7.788   & 6.829524 \\
6 & 9.688074   & 9.6875   & 9.688   & 9.65118441 \\
7 & 9.813473   & 11.7375  & 11.738  & 9.88346525 \\
\hline
\end{tabular}
\end{table}

\section{Conclusion}\label{SecV}
We have successfully modeled a discretized QAHO using a 3-qubit setup on IBM's quantum computing platform. This approach enables a detailed examination of the quantum particle's behavior over time, particularly the evolution of its wave function under the influence of an anharmonic potential. The simulation outcomes reveal time-dependent oscillations in the probability amplitudes across different computational basis states, capturing the complex, non-linear dynamics that characterize anharmonic quantum systems. The quantum circuit framework developed here provides a flexible and scalable means of implementing the unitary evolution operator associated with the anharmonic potential. Moreover, the same methodology can be extended to multi-dimensional systems by executing $N$ independent $n$-qubit circuits in parallel, each representing a single dimension, thereby improving efficiency and scalability for modeling higher-dimensional dynamics. In addition to the time-evolution simulations, three comparative analyses were performed between the VQE and exact diagonalization, perturbation theory, and the WKB approximation for states $n=0$ through $n=7$. When benchmarked against exact diagonalization, VQE slightly overestimates the energy eigenvalues, with the largest deviation being approximately 0.5 energy units at $n=7$, though both methods show consistent growth from 0 to 10 energy units. The associated error metrics confirm VQE's excellent accuracy: a MAPE of $1.1063\%$, an RMSE of $7.57\times 10^{-2}$, and an $R^2$ of 0.9999, indicating near-perfect correlation and reliability.
The comparison with perturbation theory reveals a systematic overestimation by VQE, with deviations increasing up to about one energy unit at $n=7$, corresponding to values rising from 0 to 10 (VQE) and from 0 to 11 (perturbation). The error metrics suggest moderate accuracy, with a MAPE of $6.71\%$, MSE of 0.6486, and $R^2$ of 0.6322. These values highlight strong correlation but also scalability challenges at higher energy levels, likely due to limitations of the ansatz or optimization routine. For the WKB approximation, VQE again shows systematic overestimation, with a maximum difference of roughly 1.5 energy units at $n=7$, as VQE and WKB estimates rise from 0 to 10 and from 0 to 8.5, respectively. The error metrics demonstrate good agreement overall, with a MAPE of $5.36\%$, MSE of 0.4069, RMSE of 0.638, and $R^2$ of 0.9517, confirming that while VQE follows the general trend closely, deviations become more significant at higher states. Overall, VQE excels against exact diagonalization with near-perfect accuracy, performs well against WKB, and is reasonably accurate against perturbation theory, though the latter two comparisons indicate the need for refinement at higher excitations. These results demonstrate that VQE is a powerful and reliable method for estimating quantum state energies, with clear strengths in precision and adaptability. Nevertheless, the scalability challenges identified suggest that future research should focus on optimizing the variational ansatz, improving classical optimization strategies, and developing correction techniques to further enhance accuracy and extend applicability to more complex, higher-dimensional quantum systems.
\bibliography{QHarmonic1}
\end{document}